\documentclass[a4paper,11pt]{article}
\usepackage{jcappub}
\usepackage[T1]{fontenc}
\usepackage{epstopdf}
\usepackage{float}
\usepackage{hyperref}

\title{Anisotropic strange stars in Einstein Gauss-Bonnet Gravity with Finch-Skea metric}

\author[a]{Sagar Dey,}
\author[b,1]{and  Bikash Chandra Paul. \note{Corresponding author}}

\affiliation[a]{Department of Physics, University of North Bengal, West Bengal 734014, India.}
\affiliation[b]{IUCAA Centre for Astronomy Research and Development (ICARD) and Department of Physics, University of North Bengal, West Bengal 734014, India.}

\emailAdd{sagardey231@gmail.com}
\emailAdd{bcpaul@associates.iucaa.in}

\abstract{We obtain a class of new anisotropic relativistic  solution in Einstein Gauss-Bonnet (EGB) gravity with  Finch-Skea metric in hydrostatic equilibrium. 
The relativistic solutions are employed to construct  anisotropic stellar models  for strange star with  MIT Bag equation of state $ p_{r}= \frac{1}{3} \left( \rho - 4 B_{g}\right)$, where $B_{g}$ is the Bag constants.
Considering the mass and radius of a known star PSR J0348+0432 we construct stellar models in the framework of higher dimensions. We also predict the mass and radius of  stars for  different model parameters.
The Gauss-Bonnet coupling term ($\alpha$) plays an important role in determining the density, pressure, anisotropy profiles and other features. The stability of the stellar models are probed analyzing  the different energy conditions, variation of sound speed and adiabatic stability conditions inside the star.  The central density and pressure of a star in EGB gravity are found to have higher values compared to that one obtains in Einstein gravity ($\alpha =0$). We also explore the effect of extra dimensions for the physical features of a compact object. For this we consider $D=5$ and $D=6$ to obtain a realistic stellar model and found that in the formal case both  positive and negative values of $\alpha$ are allowed. But in the later case, only $\alpha <0$ permits compact object in the Finch-Skea metric.  We determine the best fit values of the model parameters for  a number of observed stars  for acceptable stellar models. }


\begin{document}

  \maketitle
\flushbottom
\section{Introduction}\label{sec1}

The astronomical observations in recent times predict that the Einstein's general theory of relativity needs modification to describe the early and late universe. It is also felt that similar modifications are required for describing astrophysical objects namely black holes (BHs) \cite{EB,LB}. The current observations in astronomy included the gravitational waves that originated as a result of inspiraling  and finally merging of BH binaries. There is a exciting development from the observational evidences of the first image of a BH shadow by the Event Horizon Telescope \cite{KA1}. In all the cases it raises an important issue to scrutinize for an alternatives to the standard General Relativity (GR). In this direction the extended theory of gravity is considered recently to construct cosmological model of the observed universe in addition to resolve the non-renormalizabilty problem in GR.\\
The origin of the concept of higher dimensions goes back to the independent work by Kaluza \cite{kal} and Klein \cite{kle}, who tried to unify gravity with the electromagnetic interaction introducing an extra dimension. At a very high energy scale particle physics requires dimensions more than four for its consistent formulations in string theory. The Kaluza - Klein approach did not work well. The success of superstring theory led a revival of higher dimensional theories again. Eddington \cite{eddi} considered a higher dimensional spacetime to describe astrophysical objects in the usual four dimensional spacetime embedded in a flat higher dimensional spacetime. 
The issue of dimension variability was studied by 
Mandelbrot \cite{mandel} and  explained how  a ball of thin thread varies when the scale of an observer changes. In this case an object which seems to be a point object observed from a very far distance appears  a three-dimensional ball when looked at a closer distance. Consequently  as the viewer walks down the scales, the ball appears to change the form. In the case the  ball's embedding dimensions haven't changed, but the effective dimension of the contents has changed. Thus it is important to generalize the results of the usual four dimensions in the framework of higher dimensions. A number of theories so far have been  developed in higher dimensions, namely Brane-world  \cite{brane1,brane2} and non-linear Lovelock gravity \cite{love1,love2} to address different issue in cosmology and astrophysics. In the superstring theories (SST), the Gauss-Bonnet (GB) term arises as a small slope expansion in the low energy limit and the SST however is consistent in higher dimensions.
The GB combination for the quadratic terms in curvature is considered in the Einstein-Hilbert action for a proper modification of GR with ghost free. 
Zweibach \cite{zw} suggested that the string corrections due to Einstein action up to first order in the slope parameter $\alpha$ and fourth power of momenta is given by $\alpha \; (GB)$. Thereafter it was realized that the field redefinition theorem of 't Hooft and Veltman \cite{velt} can be applied in this case.  It is known that on the Einstein shell $\mathcal{R}_{\mu \nu}=0$, an action of the form $\mathcal{R}+a \; \mathcal{R}_{\mu\nu}^2 +b \; \mathcal{R}^2$ can be transformed into Ricci scalar: $\mathcal{R}$ itself (neglecting the higher order terms) by field redefinition :
\begin{equation}
\label{eqn1}
g_{\mu \nu}' = g_{\mu \nu} + a \; \mathcal{R}_{\mu \nu}+ g_{\mu \nu} \frac{a + 2b}{2-D} \; \mathcal{R}
\end{equation}
where $D$ represents number of spacetime dimensions. Later Deser and co-workers \cite{deser,boul1,boul} shown that  the actions $\mathcal{R}+ \alpha \;(GB)$ and $\mathcal{R}+ \alpha \mathcal{R}_{\mu \nu\alpha \beta}^2$ are not different. This result generalizes to all higher-order ghost terms. In cosmology an action in which the coefficients of $\mathcal{R}$ and $\alpha \;(GB)$ terms are considered arbitrary to begin with, which however determined when the relativistic cosmological solutions are tailored to fit the expected cosmological scenario. It is reported that  the signature of the coefficient of $GB$ for a realistic cosmology is  opposite to that in a superstring theory \cite{bcp}. The higher order theories of gravity have attracted much attention, as an alternative theory of gravity beyond GR, which shows quite different features from that in four dimensions. The approach has opened up a new window for several novel predictions in cosmology, astrophysics and particle physics. Einstein's gravity with a polynomial functions of Riemann tensor, Ricci tensors and Ricci scalar \cite{lancoz} is one of them. The EGB theory is a natural extension of GR to higher dimensions, which arises from the incorporation of an additional term to the standard Einstein-Hilbert action. In standard 4D, EGB and Einstein gravity are indistinguishable. The departure from the standard 4D Einstein gravity is found in more than the usual 4D. There have been many interesting results in the 5-D EGB theory ranging from the vacuum exterior solution due to Boulware and Deser \cite{boul}, which is a generalization of the Kerr-Schild vacuum solution. Thus Gauss-Bonnet terms have a rich structure in the theory which are relevant both in higher dimensional astrophysics and cosmology.\\

Consequently, a considerable research activities in understanding the astrophysical objects in higher dimensions that are known in four dimensions in modified gravity are found in the literature. The study of Vaidya radiating black-holes in EGB gravity revealed that the location of the horizons is changed considerably as compared to that in the standard 4-D gravity \cite{SG}. Wheeler \cite{wheeler}, Torii and Maeda \cite{TM} and Myers and Simons \cite{MS} obtained black hole solutions in EGB theory. Dadhich $et. \, al.$ \cite{dadhich} shown that a constant density Schwarzschild interior solution is universal, which is true both in higher dimensions Einstein theory and  EGB gravity. The Buchdahl inequality for static spheres in the usual four dimensions is generalized in five dimensional (henceforth 5D) EGB gravity \cite{buchdahl}. It is shown that the signature of the Gauss-Bonnet coupling parameter is important which plays a crucial role on the mass-to-radius ratio. It is shown that one could pack more mass in a given radius of the spherical star in five dimensional EGB gravity compared to a standard four dimensional Einstein gravity with stable configuration \cite{Goswami}. A classical isothermal sphere is generalized to a 5D EGB gravity, it exhibits a linear barotropic equation of state similar to a 4D stellar model \cite{Hansraj,Maharaj}. It is known that EGB gravity is playing an important role in obtaining relativistic solutions for constructing realistic stellar models in higher dimensional spacetime. Bhar $et.al.$ \cite{Bhar1} obtained relativistic solutions with Krori-Barua ansatz in EGB gravity and compared the physical properties of stellar models in EGB and that in GR. A static charged anisotropic fluid sphere described by Krori-Barua metric is also studied \cite{Bhar2} considering a coupled Einstein-Maxwell-Gauss-Bonnet field equations with a linear equation of state (EoS) different from MIT Bag model. Recently, Hansraj and Mkhize \cite{Hansraj1} obtained exact solutions for constructing stellar models in the Einstein-Gauss-Bonnet gravity in a six-dimensional spacetime. Considering Finch-Skea geometry, a new class of interior solutions of compact objects in five dimensional Einstein Gauss-Bonnet (EGB) gravity is obtained \cite{ift} with a linear equation of state (EoS) which permits compact objects.
 Malaver and Kasmaei \cite{MK} also constructed stellar models for anisotropic compact stars within the framework of a 5-dimensional Einstein-Gauss-Bonnet (EGB) gravity considering a linear barotropic EoS and a quadratic EoS in the framework of a generalized metric potential proposed by Thirukkanesh and Ragel \cite{tr}.\\

In the usual four dimension, Finch-Skea (henceforth, FS) metric was proposed to correct the Dourah and Ray \cite{DR} metric which was not suitable for a description of the compact objects. In GR, FS \cite{FS} metric in $D=4$ dimensions admits isotropic stellar models. Recently, FS metric was extended in a higher dimensional GR \cite{dey1} and obtained a new result that in higher dimensions FS metric accommodates anisotropic stars which is not permissible in the usual 4 dimensions. It is also shown that a charged isotropic as well as charged anisotropic stars can be accommodated in a higher dimensional GR with FS metric \cite{dey2}. \\

In compact objects the pressure in the radial and tangential directions may be different, thus anisotropic pressure in a spherically symmetric star results. Ruderman \cite{Ruderman} shown that at a high density ($> 10^{15}$ $gm/cm^3$) nuclear matter may be treated relativistically which exhibits the property of anisotropy. The reason for incorporating anisotropy is due to the fact that in the high density regime of compact stars the radial pressure ($p_r$) and the transverse pressure ($p_t$) are not equal which was pointed out by Canuto \cite{Canuto}. There are other reasons to assume anisotropy in compact stars which might occur in astrophysical objects namely, viscosity, phase transition, pion condensation, the presence of strong electromagnetic field, the existence of a solid core or type 3A super fluid, the slow rotation of fluids. The motivation of the paper is to obtain compact anisotropic strange stars with the relativistic solutions in a higher dimensional EGB gravity considering the geometry of the star described by Finch-Skea (FS) metric \cite{FS}. \\

Recent observational data for some compact stars is not understood with normal matter, consequently it is assumed that the stars are made of quarks. The possible existence of the astrophysical objects entirely made of deconfined up ($u$), down ($d$) and strange ($s$) quarks have been considered in the literature \cite{SS1,SS2,SS3} that led to a new class of stars known as strange stars. In such cases the strange quark matter hypothesis \cite{SQ1,SQ2,SQ3} is applicable, the absolute ground state for the confined state of strongly interacting matter is the strange quark matter (SQM) \cite{SS1,SS2} important for describing the compact objects. The discovery of massive compact stars with $M \sim 2 M_{\odot}$ also advocates the SQM hypothesis. Based on the recent observational data recorded by the new generation $X$-ray and $\gamma$-ray satellites, it is demonstrated \cite{XG1,XG2} that the compact objects associated with the $X$-ray burster 4U1820-30 and $X$-ray pulsar Her X1 are consistent with the strange matter equation of state (EoS). Using Shapiro delay method, Demorest and his collaborators \cite{XG3} found the mass of PSR J1614 + 2230 and predicted matter inside the star which is also consistent with the SQM. Gangopadhyay $\it et\; al.$ \cite{Gango} taken 12 known stars which are possibly strange stars (SS) candidates and estimated their radii. In those cases the MIT Bag model EoS is considered to describe the strange stars. In the literatures the MIT bag model is considered for constructing the stellar models of the strange stars. Farhi \cite{SS2} and Alcock \cite{Alcock} obtained stable models of star with SQM for the Bag constant lying between $(55-75) $ $MeV/fm^{3}$ in GR. Experimental values of the Bag parameter found from CERN-SPS and RHIC \cite{CERN} allow a wide range of values of the Bag parameter ($B_g$) consistent with the stellar models that are probed. Abbas et al. \cite{Abbas} using MIT bag model obtained anisotropic charged strange stars in $f (T )$ modified gravity in four dimensions using the diagonal tetrad fields of static space-time. Compact charged strange star model is also obtained in $f(R,T)$ theory of gravity \cite{Deb}.


The motivation of the paper is to construct stellar models of compact objects using the relativistic solutions in EGB gravity with Finch-Skea geometry. We test the stellar models in hydrostatic equilibrium in higher dimensions with the MIT Bag model. The FS metric is a modified form of Duroh-Roy metric , which is found to work well in describing compact objects. As the concept of higher dimensions is important for a consistent description of particle physics, it is legitimate to assume that the compact object is embedded in a higher dimensional spacetime which will be taken up here. The effects of extra dimensions in understanding realistic strange stars in EGB gravity will be explored following the prescription as suggested by Delgaty and Lake \cite{delg}. The stellar models will be constructed making use of the analytic solutions that satisfy all the necessary conditions for a realistic star \cite{delg}.\\

The paper is presented as follows: In Sec. $\ref{sec2}$ the basic field equations in Einstein-Gauss-Bonnet gravity in $D$-dimensions are presented for a spherically symmetric metric. In Sec. $\ref{sec3}$ we present exact solutions of the EGB equations considering Finch-Skea geometry in higher dimensions. In Sec. $\ref{sec4}$, the different criteria for a physically realistic stellar models are enumerated. The physical features of the compact stars are analyzed numerically in Sec. $\ref{sec5}$. In Sec. $\ref{sec6}$, stellar models are constructed for a known star and thereafter the model parameters are determined in different spacetime dimensions for superdense stars having  different compactness factor. Finally in Sec. $\ref{sec7}$, we discuss the results obtained in EGB gravity for realistic compact stars with concluding remarks.\\

\section{Einstein-Gauss-Bonnet gravity (EGB) gravitational action and the field equation}\label{sec2}

We consider the gravitational action with Gauss-Bonnet terms is given by,
\begin{equation}
 \label{eqn2}
 S = \int \sqrt{-g}\left[\frac{1}{2}(\mathcal{R}+\alpha\;L_{GB} )\right] d^{D}x + S_{m}.
\end{equation}
where, $\mathcal{R}$ is the Ricci scalar, $L_{GB}$ is the Gauss-Bonnet terms, $S_{m}$ is the matter action, $g$ is the metric term in higher dimensions and the dimensional coupling parameter $\alpha$ . The Gauss-Bonnet Lagrangian is given by, $L_{GB} = \mathcal{R}^{2}-4\mathcal{R}_{cd}\mathcal{R}^{cd} + \mathcal{R}_{abcd}\mathcal{R}^{abcd}$, where the indices $a$, $b$, $c$ and $d$ are indices (0,...., $D-1$).
The field equation in the presence of matter field is derived from Eq. (\ref{eqn2}) which is given by, \\
\begin{equation}
\label{eqn3}
 G_{ab}+\alpha H_{ab} = 8\pi T_{ab}.
\end{equation}
where $G_{ab}$ denotes the Einstein tensor, $T_{ab}$ is the Total energy-momentum tensor and the Lanczos tensor is given by,
\begin{equation}
\label{eqn4}
 H_{ab}= 2(\mathcal{R} \mathcal{R}_{ab}-2 \mathcal{R}_{ac}\mathcal{R}^{c}_{b}-2\mathcal{R}^{cd}\mathcal{R}_{acbd}+\mathcal{R}^{cde}_{a}\mathcal{R}_{bcde})-\frac{1}{2}g_{ab}L_{GB}.
\end{equation}
For a spherically symmetric spacetime the $D$-dimensional line element is given by,\\
\begin{equation}
\label{eqn5}
 ds^{2}= - e^{2\nu(r)}dt^{2}+e^{2\lambda(r)}dr^{2}+r^{2} d\Omega^{2}_{D-2}.
 \end{equation}
 where, $d\Omega^{2}_{D-2}$ is metric on a unit $(D-2)$ -sphere and $\nu(r)$, $\lambda(r)$ are the metric potentials. The energy momentum tensor of the matter is given by
 \begin{equation}
 \label{eqn6}
 T^{a}_{b}=diag (-\rho, p_{r}, p_{t}, .........,p_{t}).
 \end{equation}
 for anisotropic matter distribution in $D$- dimensions, $\rho$ is the matter density, $p_{r}$ is the radial pressure and $p_{t}$ is the tangential pressure. The energy-momentum conservation law $T^{ab}\;_{;\;b}=0$ in $D$-dimensions is given by,
 \begin{equation}
 \label{eqn7}
 \frac{dp_{r}}{dr}+(\rho +p_{r} )\;\nu'(r) + \frac{(D-2)(p_{t}-p_{r})}{r}=0.
 \end{equation}
 where $()'$ represents derivative with respect to radial coordinate $r$. Using the metric Eq. ($\ref{eqn5}$) in the field Eq.($\ref{eqn3}$) we determine the components of D-dimensional Einstein-Gauss-Bonnet (EGB) field equation which are given by,
\begin{equation}
\label{eqn8}
 \rho =\frac{(D-2) e^{-2\lambda}}{2 r^{2}} \left[2r\lambda'(r)+(D-3)(e^{2\lambda}-1)+\frac{ \alpha f(\lambda) (4r \lambda'(r)+(D-5) (e^{2\lambda}-1))} { r^{2}}\right].
\end{equation}
\begin{equation}
\label{eqn9}
 p_{r}= \frac{(D-2) e^{-2\lambda}}{2 r^{2}} \left [ 2 r\nu'(r)-(D-3)(e^{2\lambda}-1) + \frac{\alpha f(\lambda) (4r \nu'(r)-(D-5) (e^{2\lambda}-1))} { r^{2}}\right].
\end{equation}
 where $f (\lambda)= (1- e^{-2\lambda})$ and for simplicity we have taken $8\pi G_{D} = c^{2}=1$.
 The non-linear equations in EGB-gravity given by  Eqs.($\ref{eqn7}$)-($\ref{eqn9}$) are used to obtain stellar models for compact stars. The stellar models are explored numerically of which GR limit corresponds to $\alpha = 0$. The the effects of the coupling parameters for a realistic stellar models are studied in the  next section.

\section{Relativistic solutions in EGB gravity}\label{sec3}

 We consider a higher dimensional Finch-Skea (FS) metric \cite{FS}  given by 
\begin{equation}
\label{eqn11}
e^{2\lambda(r)}=\left(1+\frac{r^{2}}{R^{2}}\right).
\end{equation}
where, $R$ is a dimensional constant, prescribing a specific geometry of the interior spacetime. The parameter $R$ can be determined using the matching conditions of the metric with the exterior vacuum solution at the boundary. For a strange star configuration, we consider the strange quark matter distribution described by a simplest phenomenological MIT Bag model.
Considering non-interacting quarks with all the three flavours confined in a bag, massless in nature, the quark pressure of the strange Quark Matter (SQM) is given by,
\begin{equation}
\label{eqn12}
 p_{r}=\sum_{f=u,d,s} p^{f} - B_{g}.
\end{equation}
where $p^{f}$ is the individual pressure of the quarks, $B_{g}$  the total external bag pressure which is a constant quantity. For free quarks the relation between the individual quark pressure $p^{f}$ and energy density of the individual quark flavour can be written as, $p^{f}=\frac{1}{3}\rho^{f}$. Thus inside the bag the deconfined quarks have the total energy density $\rho$ given by
\begin{equation}
\label{eqn13}
 \rho =\sum_{f=u,d,s} \rho^{f}+ B_{g}.
\end{equation}
Using Eqs.(\ref{eqn12}) and (\ref{eqn13}) one gets the MIT bag model EoS which is given by
\begin{equation}
\label{eqn14}
 p_{r}= \beta\left( \rho - 4 B_{g}\right).
\end{equation}
where $\beta$ is a constant. The numerical value of $\beta$ is  0.28 for massive strange quarks having mass 250 MeV and $\frac{1}{3}$ for massless strange quarks.
Thus the EoS for massless SQM is
\begin{equation}
\label{eqn15}
 p_{r}= \frac{1}{3} \left( \rho - 4 B_{g}\right).
\end{equation}
Farhi and Jaffe \cite{SS1} shown that for  non-interacting massless quarks SQM might give the true ground state for a strongly interacting matter distribution when $B_{g}$ lies between $57 \; MeV/fm^{3}$ and $94 \;MeV/fm^{3}$. \\
In EGB gravity with FS geometry,  there are five unknowns in the stellar models, namely, $R$, $r$, $\alpha$, $D$, $B_{g}$ and we have only three equations (two matching conditions and radial pressure vanishes, $p_r(r=b)=0$ at the boundary), therefore two more $ad \, hoc$ assumptions can be made for a physically realistic stellar model. Therefore stellar models can be constructed for known values of the model parameters and can be tested for a realistic model. Now using Eqs.($\ref{eqn8}$) and ($\ref{eqn9}$) in MIT Bag model EoS given by Eq. (\ref{eqn15}), we determine the metric component $\nu (r)$, which is given by
\begin{equation}\label{eqn16}
\frac{d\nu(r)}{dr}= \frac{r \left(-\frac{4 B_{g} \bar{R}^{4}}{3 (D-2)}+Y+\frac{1}{3} \left(Y+\frac{2
   \alpha  R^2}{\bar{R}^{2}}+R^2\right)\right)}{R^2 \left(2 \alpha +\bar{R}^{2}\right)}.
\end{equation}
where, $Y= \frac{1}{2} \alpha  (D-5)+\frac{1}{2} (D-3) \bar{R}^{2}$.\\
On integration once we get,
\[
\nu (r)= \nu_{0}+ \frac{1}{6} \log \bar{R}^{2}
\]
  \begin{equation}\label{eqn17}
  +\frac{\left(2 \alpha +\bar{R}^{2}\right) \left(B_{g} \left(6 \alpha - \bar{R}^{2}\right)+ D^2- 5 D + 6\right)- \alpha  \left(8 \alpha B_{g}+D^2-3
   D+2\right) \log \left(2 \alpha +\bar{R}^{2}\right)}{3(D-2) R^2}.
 \end{equation}
where $\nu_0$ is an integration constant and we denote $r^{2}+R^{2}= \bar{R}^{2}$ and
The radial pressure of a star at the boundary ($r=b$) is $p_{b}$ = 0, which is another constraint to satisfy. The Bag constant $B_{g}$ now can be determined in terms of the density at the boundary ($\rho_{b}$) which is given by,
\begin{equation}
\label{eqn18}
 B_{g}= \frac{1}{4} \rho_{b}.
\end{equation}
For a given value of  $B_{g}$, the metric parameter ($\nu $) can be determined from Eq.(\ref{eqn17}). We construct stellar models fixing the Bag constant at $B_{g}= 60 MeV/fm^{3}$.
The energy density and pressures are given by 
\begin{equation}
\label{eqn19}
 \rho = \frac{2 \alpha(D-2) R^2}{\bar{R}^{6}}+\frac{\alpha (D-5) (D-2)}{2 \bar{R}^{4}}+\frac{(D-2) R^2}{\bar{R}^{4}}+\frac{(D-3)
 (D-2)}{2 \bar{R}^{2}}.
\end{equation}
\[
p_{r}= \frac{Y_{1} \left(b^4 \left((D-3) r^4+r^2 \left(\alpha(D-5)+2 (D-2) R^2\right)+(D-1) R^2 \left(\alpha +R^2\right)\right)\right)}{6 \left(b^2+R^2\right)^{3} \bar{R}^{6}}
\]
\[
+  \frac{Y_{1} b^2 \left(r^4 \left(\alpha 
   (D-5)+2(D-2)R^2\right)+4 r^2 R^2 \left(\alpha (D-4)+(D-1)R^2\right)+R^4 Y_{2}\right)}{6 \left(b^2+R^2\right)^{3} \bar{R}^{6}}
\]
\begin{equation}
\label{eqn20}
+  \frac{Y_{1} R^2 \left((D-1) r^4 \left(\alpha +R^2\right)+r^2 R^2 Y_{2}+(D+1) R^4 \left(2 \alpha +R^2\right)\right)}{6 \left(b^2+R^2\right)^{3} \bar{R}^{6}}.
\end{equation}
where, $b$ is the radius of a star, $Y_{1} = (D-2) \left(b^2-r^2\right)$ and $Y_{2}= -3 \alpha +3 \alpha D +2 D R^2$. 
\section{Constraints for stellar models}\label{sec4}

The stellar models  are obtained that satisfy the following conditions \\

$\bullet$ The energy density ($\rho$), radial  pressure ($p_{r}$) and transverse pressure ($p_{t}$) are positive inside. 

$\bullet$ The radial pressure drops to zero ($ p_{r}|_{r=b} = 0$) at the boundary, 

$\bullet$ For a stable configuration, both $ v_{r}^{2} = \frac{dp_{r}}{d\rho}$ and $ v_{t}^{2} = \frac{dp_{t}}{d\rho} $ must be $\leq 1$.  

$\bullet$ The gradient of the pressures and energy-density  inside  should satisfy  $\frac {dp_{r}}{dr} < 0 $, $\frac {dp_{t}}{dr} < 0 $, $ \frac {d\rho} {dr} < 0 $.

$\bullet$ $\textbf{Boundary condition in EGB gravity}:$ The interior metric can be matched with a definite exterior metric that leads to a vacuum solution. In the present case we consider  EGB gravity, thus the exterior solution obtained by Wiltshire in $D$-dimensions   will be matched at the boundary \cite{boul} as follows:
\[
 ds^{2}= - F(r)dt^{2}+\frac{dr^{2}}{F(r)}+r^{2} d\Omega^{2}_{D-2}.
\]
where, $F(r)= 1+ \frac{r^{2}}{8\bar{\alpha}}\left(1-\sqrt{1+\frac{32 \bar{\alpha} G M}{r^{D-1}}}\right)$, with a constant $\alpha$ being the string tension and $M$ the gravitational mass. Thus the matching conditions at the boundary $r=b$ with metric given by eq. (\ref{eqn5}) are
\begin{equation}
\label{eqn22}
e^{2\nu}|_{ (r=b)} = F(b), \;\;\;\; \; \; e^{2\lambda} |_{(r=b) } =\frac{1}{F(b)}.
\end{equation}

\section{Physical features of the model}\label{sec5}

The relativistic solutions in EGB gravity obtained here will be employed to construct stellar models satisfying all the criteria of a physically realistic solution. To begin with we consider a compact object namely, PSR J0348+0432 with its observed mass $M= 2.01\pm$ 0.04 $M_{\odot}$ \cite{Anto} which is precisely measured by observations.

\subsection{\bf{Energy-density and Pressure}}
The radial variation of the energy density ($\rho$)and the radial pressure ($p_{r}$) of PSR J0348+0432  are plotted in Figs.(1) and (2) respectively for both positive and negative  $\alpha$ in $5$- dimensions. It is evident that as  the coupling parameter $\alpha$ decreases, the radial pressure at the centre decreases.  The radial pressure inside the star is found more for positive  $\alpha$ compared to that with negative $\alpha$.

\begin{figure}[H]
 \centering
 \includegraphics[width=0.5\textwidth]{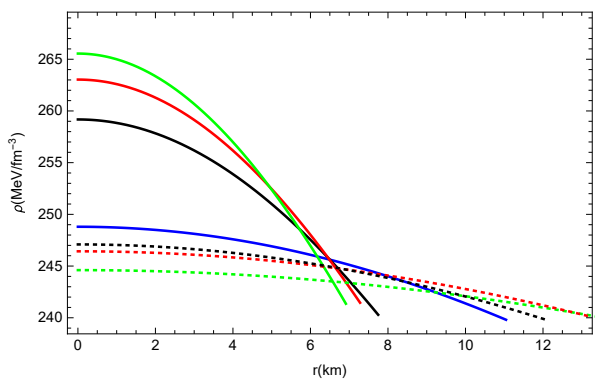}
 \caption {Variation of $\rho$ in PSR J0348+0432 for different $\alpha$ values taking $B_{g}= 60 MeV/fm^{3}$ ($\alpha= 0$ (Blue) , $\alpha= 0.5$ (Black), $\alpha= 1.0$ (Red), $\alpha= 3.0$ (Green), $\alpha= - 0.5$ (Black, Dotted), $\alpha=- 1.0$ (Red, Dotted) and $\alpha=- 3.0$(Green, Dotted)).}
\end{figure}

\begin{figure}[H]
 \centering
 \includegraphics[width=0.5\textwidth]{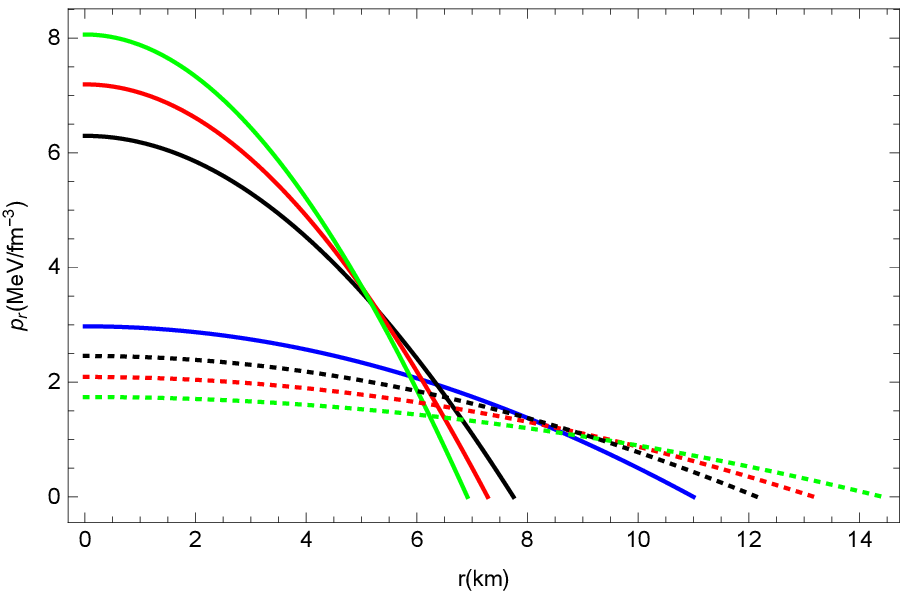}
 \caption {Variation of $\rho$ in PSR J0348+0432 for different $\alpha$ values taking $B_{g}= 60 MeV/fm^{3}$ ($\alpha= 0$ (Blue) , $\alpha= 0.5$ (Black), $\alpha= 1.0$ (Red), $\alpha= 3.0$ (Green), $\alpha= - 0.5$ (Black, Dotted), $\alpha=- 1.0$ (Red, Dotted) and $\alpha=- 3.0$(Green, Dotted)).}
\end{figure}

\subsection{\bf{Analysis of Anisotropy in the Compact objects }}
We define anisotropic factor as,
\begin{equation}
\label{eqn23}
 \Delta = p_{t}- p_{r}.
\end{equation}
which varies with the radius ($r$). In Fig.(3), the radial variation of anisotropy is plotted for different values of $\alpha$ in $D=5$ dimensions, we get  $p_{t} > p_{r}$.\\

\begin{figure}[H]
 \centering
 \includegraphics[width=0.5\textwidth]{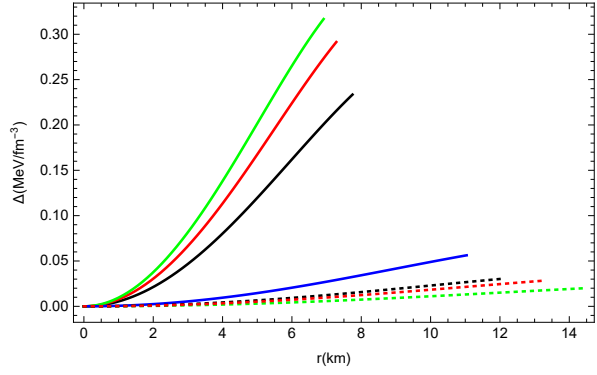}
 \caption {Radial variation of anisotropic stress factor in PSR J0348+0432 for different $\alpha$ values taking $B_{g}= 60 MeV/fm^{3}$ ($\alpha= 0$ (Blue) , $\alpha= 0.5$ (Black), $\alpha= 1.0$ (Red), $\alpha= 3.0$(Green), $\alpha= - 0.5$ (Black, Dotted), $\alpha=- 1.0$(Red, Dotted) and $\alpha=- 3.0$ (Green, Dotted)).}
\end{figure}

\subsection{\bf{Stability of the Stellar Models}}
In this section,  the stability of the stellar models are tested making use of the two methods given by  (i) Herrera's cracking concept and  determining the (ii) the adiabatic index.

\subsubsection{\bf{Study of Herrera's cracking concept}}
For a physically acceptable stellar model the radial sound speed ($v_{r}^{2}=\frac{dp_{r}}{d\rho}$) and that for transverse ($v_{t}^{2}=\frac{dp_{t}}{d\rho}$) should satisfy, $i.e.$, $0< v_{r}^{2} \leq 1$ and $0< v_{t}^{2} \leq 1$ \cite{HH}. Using Eq.($\ref{eqn15}$) we get
\[
v_{r}^{2}=\frac{dp_r}{dr}= \frac{1}{3}.
\]
which signals that the fluid inside is causal. Here $v_{r}^{2} <1$ always allowing a star which is radially stable. We  also plot in Fig.(4)  the radial variation of $v_{r}^{2}$ and $v_{t}^{2}$ respectively in $D=5$. The radial variation of $|v_{t}^{2}-v_{r}^{2}|$ is plotted in Fig. (5), which is always less than unity and tested for a set of values of $\alpha$, which admits  realistic stellar models.

\begin{figure}[H]
 \centering
 \includegraphics[width=0.5\textwidth]{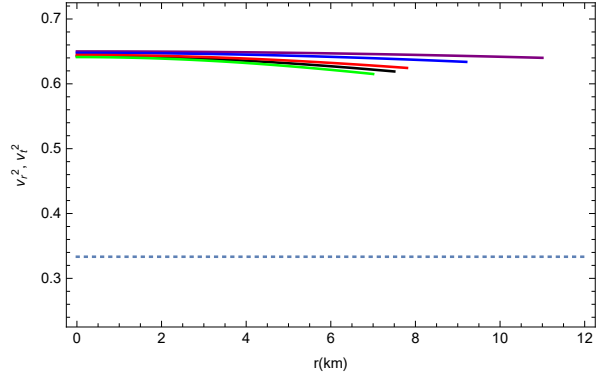}
 \caption {Radial variation of radial speed of sound $v_r^2$ ({\it dashed line}) and transverse speed of sound $v_t^2$ ({\it solid line, dotted line}) in PSR J0348+0432 for different $\alpha$ values taking $B_{g}= 60 MeV/fm^{3}$ ($\alpha= 0$ (Blue) , $\alpha= 0.5$ (Black), $\alpha= 1.0$ (Red), $\alpha= 3.0$(Green), $\alpha= - 0.5$ (Black, Dotted), $\alpha=- 1.0$(Red, Dotted) and $\alpha=- 3.0$ (Green, Dotted)).}
\end{figure}

\begin{figure}[H]
 \centering
 \includegraphics[width=0.5\textwidth]{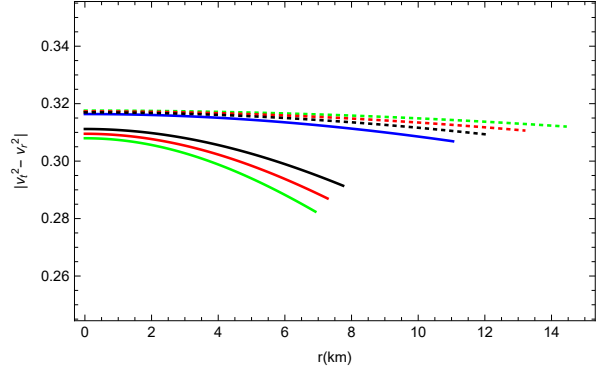}
 \caption {Radial variation of $v_{t}^{2}-v_{r}^{2}$  in PSR J0348+0432 for different $\alpha$ values taking $B_{g}= 60 MeV/fm^{3}$ ($\alpha= 0$ (Blue) , $\alpha= 0.5$ (Black), $\alpha= 1.0$ (Red), $\alpha= 3.0$(Green), $\alpha= - 0.5$ (Black, Dotted), $\alpha=- 1.0$(Red, Dotted) and $\alpha=- 3.0$ (Green, Dotted)).}
\end{figure}

\subsubsection{\bf{Study of Adiabatic index}}

The stiffness of the matter  is characterized by the adiabatic index which has significant importance in understanding the stability of a relativistic compact object. The dynamical stability against infinitesimal radial adiabatic perturbation of stellar systems was first initiated by Chandrasekhar. The stability of a star  further gets confirmed if the adiabatic index greater than $\frac{4}{3}$ as was pointed out by Bondi \cite{Bondi}. For anisotropic fluid distribution the two different adiabatic index are defined as
\begin{equation}
\label{eqn24}
 \Gamma =\frac{\rho+ p_{r}}{p_{r}} \frac{dp_{r}}{d\rho}.
\end{equation}
 $\Gamma$ in EGB gravity is highly non-linear which is to be analyzed numerically.  The radial variation of $\Gamma$ is plotted for different values of $\alpha$ in Fig.(6) and  noted  that stable models are permissible.

\begin{figure}[H]
 \centering
 \includegraphics[width=0.5\textwidth]{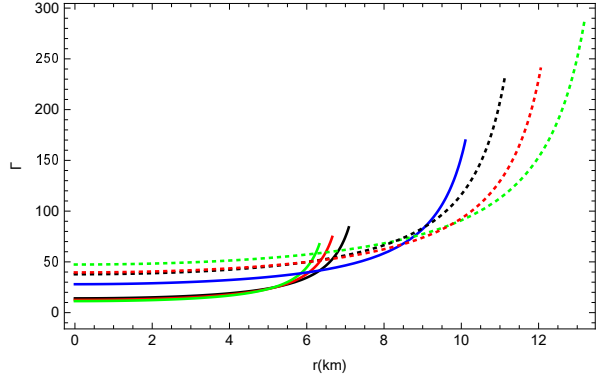}
\caption{Radial variation of $\Gamma$ for different $\alpha$ in PSR J0348+0432 taking $B_{g}= 60 MeV/fm^{3}$ in $D=5$ dimensions.($\alpha= 0$ (Blue) , $\alpha= 0.5$ (Black), $\alpha= 1.0$ (Red), $\alpha= 3.0$(Green), $\alpha= - 0.5$ (Black, Dotted), $\alpha=- 1.0$ (Red, Dotted) and $\alpha=- 3.0$ (Green, Dotted)).}
\end{figure}

\subsection{\bf{Energy conditions}}

The weak energy condition ($\rho+p_{r} \geq 0$, $\rho+p_{t} \geq 0$), strong energy condition ($\rho-p_{r}-2 p_{t} \geq 0$) and the dominant energy condition ($\rho - p_{r} \geq 0$, $\rho -p_{t} \geq 0$) are studied numerically. In Figs.(7) we draw all the energy conditions (EC) and found that EC satisfies for the set of values of the model parameters in both GR and beyond i.e.,  EGB. We study relative merits of the theories.

\begin{figure}[H]
\begin{center}$
\begin{array}{lll}
\includegraphics[width=40mm]{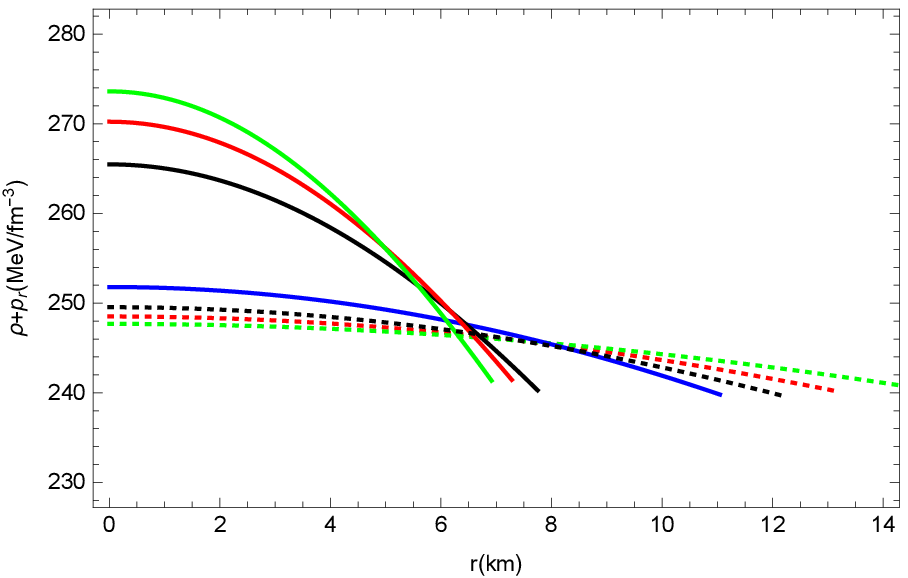}
\label{5a}&
\includegraphics[width=40mm]{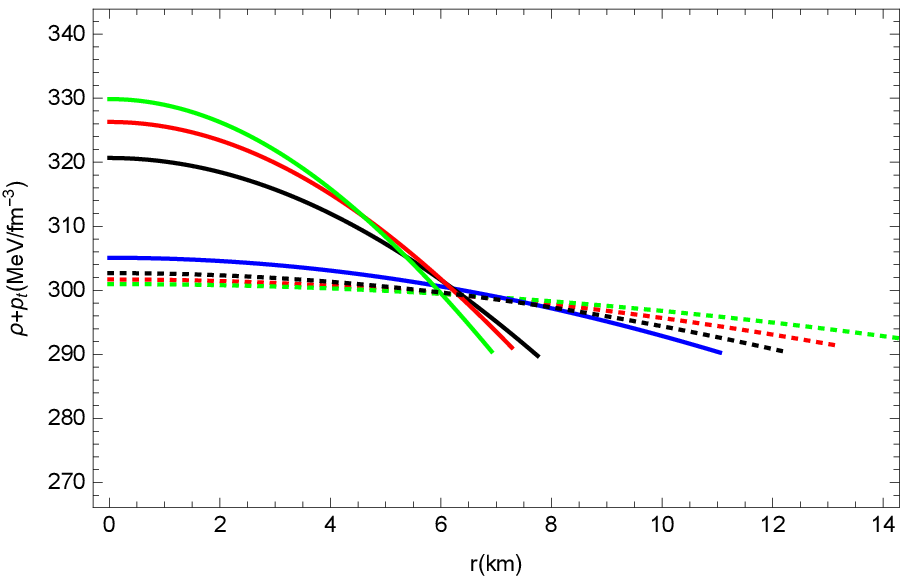}
\label{5b}&
\includegraphics[width=40mm]{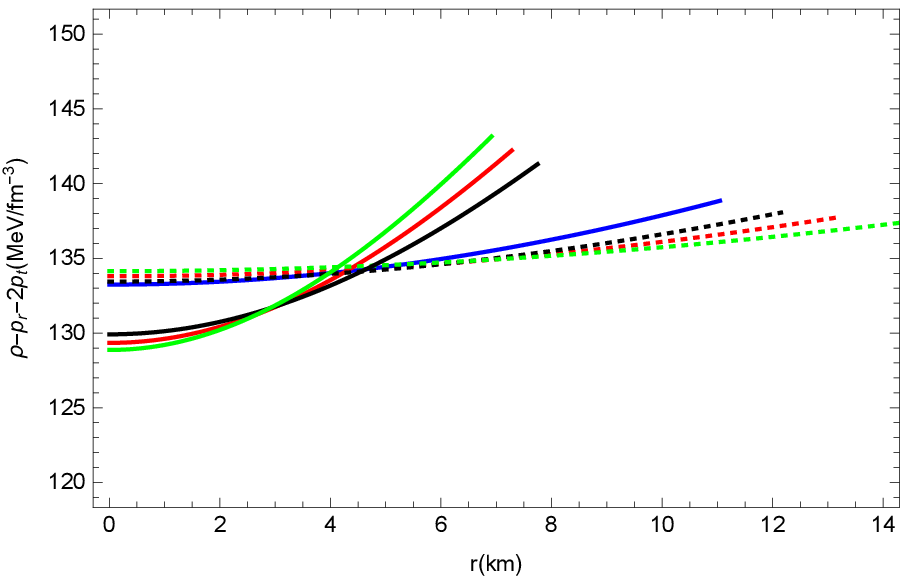}
\label{5c}
\end{array}$
\end{center}

\begin{center}$
\begin{array}{rr}
\includegraphics[width=40mm]{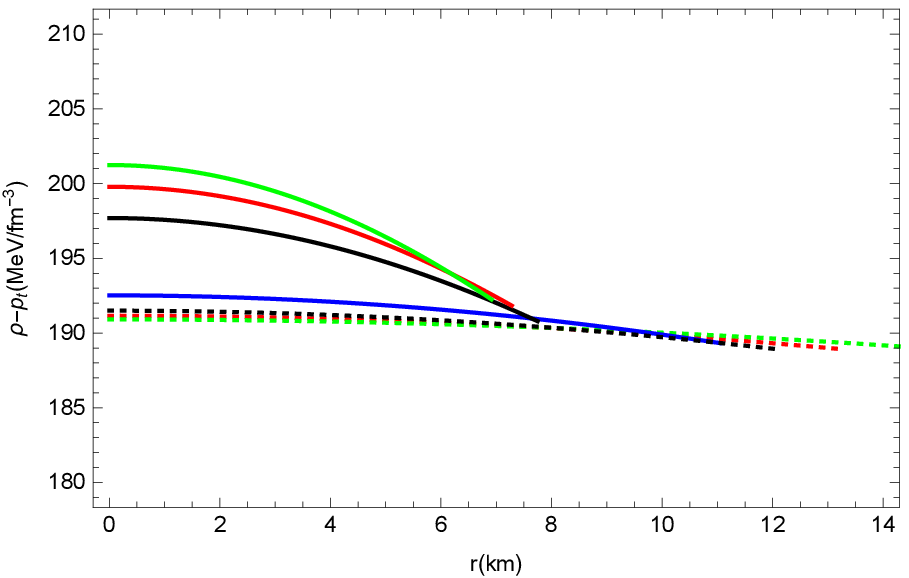}
\label{5d}&
\includegraphics[width=40mm]{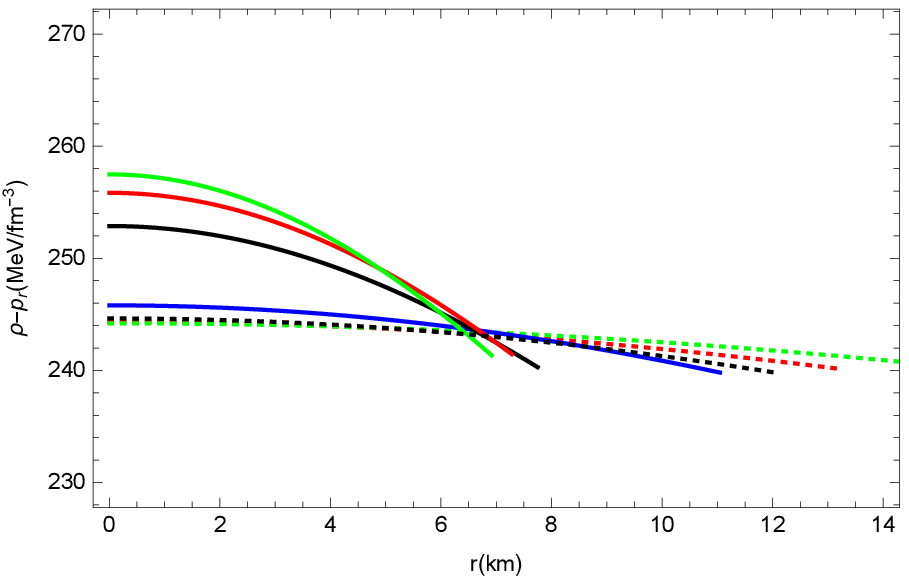}
\label{5e}
\end{array}$
\end{center}
\caption{WEC$_{r}$ (Upper left), WEC$_{t}$ (Upper Middle), SEC (Upper right), {DEC$_{r}$} (Lower left) {DEC$_{t}$} (Upper middle) for different $\alpha$ in PSR J0348+0432 taking $B_{g}= 60 MeV/fm^{3}$ in $D=5$ dimensions. ($\alpha= 0$ (Blue), $\alpha= 0.5$ (Black), $\alpha= 1.0$ (Red), $\alpha= 3.0$(Green), $\alpha= - 0.5$ (Black, Dotted), $\alpha=- 1.0$ (Red, Dotted) and $\alpha=- 3.0$ (Green, Dotted))}
\end{figure}

\subsection{\bf{Tolman-Oppenheimer-Volkoff (TOV) equation}}

In  compact object three different forces, namely gravitational force, hydrostatics force, and anisotropic force are operating. The hydrostatic equilibrium conditions inside the star in the presence of all the three forces is ensured from the study of the TOV-equation, which is given by\cite{tov}:
\begin{equation}
 \label{eqn25}
 -\frac{M_{G}(\rho+p_{r}) e^{\frac{\nu-\lambda}{2}}}{r}-\frac{dp_{r}}{dr}+\frac{(D-2)}{r}(p_{t}-p_{r})=0.
\end{equation}
where $M_{G}$ is the gravitational mass within the radius $r$ derived from the Tolman-Whittaker formula and the Einstein's field equations. The gravitational mass is given by,
\begin{equation}
 \label{eqn26}
 M_{G}=  r\;e^{\frac{\lambda-\nu}{2}} \nu'.
\end{equation}
Thus, mathematically, the equilibrium of a stellar configuration is ensured subject to the stability under the presence of the three force gradients,
\begin{equation}
 \label{eqn27}
 f_{g}+ f_{h}+ f_{a}=0.
\end{equation}
where,
\begin{equation}
 \label{eqn28}
 f_{g}= -\nu'(\rho+p_{r}),\ \ f_{h}= -\frac{dp_{r}}{dr}\ \ \textit{and}\ \ f_{a}= \frac{(D-2)}{r}(p_{t}-p_{r}).
\end{equation}
The profile of the radial variation of the three different force gradients inside the star are shown in Fig.(8) for different values of $\alpha$.

\begin{figure}[H]
 \centering
 \includegraphics[width=0.5\textwidth]{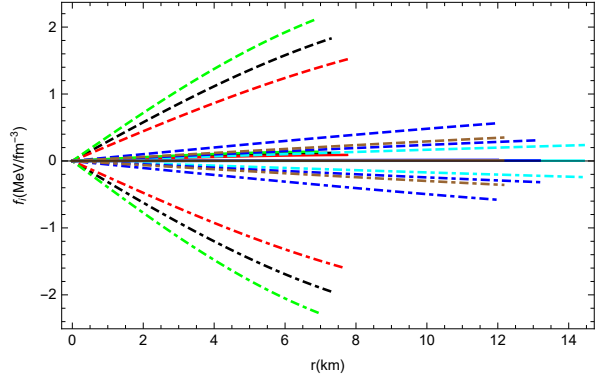}
 \caption{TOV relation in PSR J0348+0432. The blue, black, red and green line for $\alpha= 0$, $\alpha= 0.5$, $\alpha= 1.0$ and $\alpha= 3.0$, respectively for $B_{g}= 60 MeV/fm^{3}$ with different values of $R$ given in Table 1 [Here, $F_{g}$(DotDashed),$F_{a}$(Soild) and $F_{h}$(Dashed)] }
\end{figure}

\subsection{\bf{Mass-radius relationship}}

The effective mass of the stellar object within the radial distance $r$ is given by,
\begin{equation}
 \label{eqn31}
 M= \int_0^b 4\pi r^{2}\; \rho\;dr.
\end{equation}
The mass-radius variation for different $\alpha$ for $D=5$ is drawn in Fig.(9).

\begin{figure}[H]
 \centering
 \includegraphics[width=0.5\textwidth]{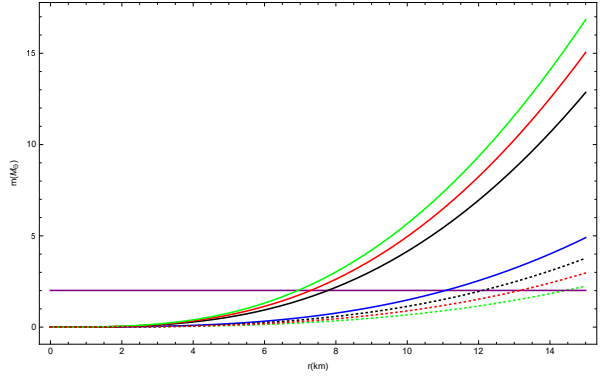}
 \caption{Mass against the radial distance $r$. The purple, blue, red, black and green line for $\alpha= 0$, $\alpha= 0.01$, $\alpha= 0.5$, $\alpha= 1.0$ and $\alpha= 3.0$,  $\alpha= - 0.5$ (Black, Dotted), $\alpha=- 1.0$(Red, Dotted) and $\alpha=- 3.0$ (Green, Dotted)respectively for $B_{g}= 60 MeV/fm^{3}$ with different values of $R$ mentioned in Table 1 }
\end{figure}

\section{Stellar models in  D=5 and  D=6 dimensions}
\label{sec6}
In order to fit the robustness of our approach in obtaining stellar model of compact objects within the framework of higher dimensional EGB gravity, the following prescription are used in the case of other compact objects with different compactification factors:  In the theory there are five unknowns, $viz.$, $R$, $r$, $\alpha$, $D$, $B_{g}$ and three equations, for this we assumed  two of the parameters in the model, namely, $D$ and $B_{g}$ to construct stellar models. In the previous sections we obtain stellar models for PSR J0348$\pm$ 0432 with mass $M= 2.01$ $\pm$0.04 $M_{\odot}$ and obtain stellar models for different values of the EGB parameters ($\alpha$). The compactification factor of a star is a dimensionless number defined as ($u= \frac{M}{b}$) where $M$ is the mass and $b$ is the radius both are expressed in the gravitational unit. We now consider stars of same mass to begin with to estimate radius that can be accommodated in our model for different $\alpha$ values. Taking the mass of the star given by $M= 2.01 \; M_{\odot}$ in both EGB and GR we can construct stellar models.
In  Table-(1),  the permissible values of the model parameters are tabulated which  accommodates stars of given radius and the corresponding compactification factor  $i.e., $ mass of the star may be different. For $\alpha = 0$ in $D=5$-dimensional Einstein gravity it permits a star of radius $b=11.05 \, km.$. In Table- (2), we tabulated the parameters in $D=6$ dimensions also for a star with  mass $ M= 2.01 \, M_{\odot}$. Considering five known compact objects, namely, Vela X-1, Cen X-3, EXO 1745-268, SAX J1748.9-2021 and M13 we  determined the best-fit model parameters for realistic stellar model, which are tabulated in Table-(3). It is evident that stellar models are permitted with both positive and negative $\alpha$ in five dimensional FS manifold for such known stars, but in $D=6$ only negative values of $\alpha$ are allowed for a stable configuration. 

\begin{table}[H]

\begin{center}
\begin{tabular}{|c|c|c|c|c|c|} \hline

$\alpha$ & $R$ & Star's Radius ($r=b$) & $\rho_{0} $ & $p_{r0} $ & $u=\frac{M}{b}$ \\
&& (km.) & (gm/cm$^{3}$) & (dyne/cm$^{2}$) & \\ \hline
-3.0& 120& 14.42 &4.38 $\times$ $10^{14}$ &2.77 $\times$ $10^{33}$ & 0.20 \\ \hline
-1.0 & 100& 13.17& 4.39 $\times$ $10^{14}$ &3.34 $\times$ $10^{33}$ &0.22\\ \hline
-0.5& 90 & 12.50& 4.40 $\times$ $10^{14}$& 3.71 $\times$ $10^{33}$& 0.23\\ \hline
0 & 70 & 11.05 & 4.45 $\times$ $10^{14}$ & 4.72 $\times$ $10^{33}$ & 0.26\\ \hline
0.5 & 34 & 7.75 & 4.97 $\times$ $10^{14}$ & 10.81 $\times$ $10^{33}$ & 0.38 \\ \hline
1.0 & 30 & 7.28 & 5.16 $\times$ $10^{14}$ & 13.33 $\times$ $10^{33}$ & 0.40 \\ \hline
3.0 & 27 & 6.91 & 5.36 $\times$ $10^{14}$ & 14.53 $\times$ $10^{33}$ & 0.42 \\ \hline
\label{Tab1}
\end{tabular}
\caption{Numerical values of the physical parameters for the strange star candidate PSR J0348+0432 having mass 2.01$\pm$0.04 $M_{\odot}$ due to the different values of $\alpha$ in $D=5$ dimensions where $R$ is the metric parameter.}
\end{center}
\end{table}

\begin{table}[H]

\begin{center}
\begin{tabular}{|c|c|c|c|c|c|} \hline
\label{Tab2}
$\alpha$ &$ R$ & Star's Radius ($r=b$) & $\rho_{0} $ & $p_{r0} $ & $u=\frac{M}{b}$ \\
&& (km.) & (gm/cm$^{3}$) & (dyne/cm$^{2}$) & \\ \hline
0 & 164.4 & 11.0& 4.32 $\times$ $10^{14}$ & 8.04 $\times$ $10^{32}$ & 0.26 \\ \hline
-10 & 78.26 & 8.2 & 4.34 $\times$ $10^{14}$ & 1.95 $\times$ $10^{33}$ & 0.36\\ \hline
-15 & 70.0 & 7.9 & 4.36 $\times$ $10^{14}$ & 2.21 $\times$ $10^{33}$ & 0.37 \\ \hline
-30 & 65.71 & 7.7 & 4.37 $\times$ $10^{14}$ & 2.35 $\times$ $10^{33}$ & 0.38\\ \hline
-40 & 54.31 & 7.2 & 4.38 $\times$ $10^{14}$ & 2.88 $\times$ $10^{33}$ & 0.41\\ \hline
-50 & 49.44 & 7.0 & 4.39 $\times$ $10^{14}$ & 3.09 $\times$ $10^{33}$ & 0.42\\ \hline
\end{tabular}
\caption{Numerical values of the physical parameters for PSR J0348+0432 having mass 2.01$\pm$0.04 $M_{\odot}$ with $\alpha <0$ in $D=6$ dimensions}
\end{center}
\end{table}

\begin{table}[H]
\begin{center}
\begin{tabular}{|c|c|c|c|c|c|c|} \hline
\label{Tab3}
Stars & Mass in M$_{\odot}$ & Radius& $\alpha$ & R & $\rho_{0} $ & $p_{r0}$ \\
&& $r=b$ $km.$ & && (gm/cm$^{3}$) & (dyne/cm$^{2}$) \\ \hline
Vela X-1 \cite{ZGG} & 1.77 $^{+0.88}_{-0.88}$ & 9.56$\pm$ 0.08 & 10 & 56 & 4.47$ \times$ 10$^{14}$& 5.67$ \times$ 10$^{33}$ \\
&& &-25 &55.3&4.46 $\times$ 10$^{14}$ & 5.68$\times$ 10$^{33}$ \\ \hline
Cen X-3 \cite{ZGG} & 1.49 $^{+0.08}_{-0.08}$ & 9.17$\pm$0.13& 1 & 56 & 4.44 $ \times$ 10$^{14}$ & 5.86 $ \times$ 10$^{33}$ \\
&& &-2 &55& 4.45 $ \times$ 10$^{14}$ & 5.14 $ \times$ 10$^{33}$\\\hline
EXO & 1.65$^{+0.21}_{-0.31}$ & 10.5$\pm$ 0.16 & 5 & 70& 4.41$ \times$ 10$^{14}$& 4.32 $ \times$ 10$^{33}$ \\
1745-268 \cite{FO} && &-20 &69.6&4.41$ \times$ 10$^{14}$ &3.47$ \times$ 10$^{33}$\\ \hline
SAX & 1.81$^{+0.25}_{-0.37}$ & 11.7$\pm$ 1.7 & 2.0 & 84& 4.38 $ \times$ 10$^{14}$ & 3.74$ \times$ 10$^{33}$ \\
J1748.9-2021\cite{FO} && &-105 &81.5& 4.39 $ \times$ 10$^{14}$ &3.82$ \times$ 10$^{33}$\\ \hline
M13 \cite{NAD} & 1.38$^{+0.08}_{-0.23}$ & 9.95$^{+0.24}_{-0.27}$ & 20 & 84& 4.42 $ \times$ 10$^{14}$& 4.03 $ \times$ 10$^{33}$ \\
&& &-18 &68&4.41$ \times$ 10$^{14}$ & 4.06$ \times$ 10$^{33}$\\\hline
\end{tabular}
\caption{Best fitted values of model parameters for different known compact object in $D=5$ dimensions }
\end{center}
\end{table}

\section{Discussion}\label{sec7}
We present  a class of new relativistic solutions in EGB gravity with Finch-Skea metric in higher dimensions. Models of compact objects in hydrostatic equilibrium are constructed considering FS-metric in $D=5$ and $D=6$ dimensions respectively and distinguished the features of the stellar models. We consider EoS of the MIT Bag model for the matter description in the strange stars to   construct the  stellar models for known mass of star for a set of permitted values of the model parameters. The stellar models obtained here satisfies all the criteria required for a stable star. As the field equations are highly non-linear we adopt numerical method to analyze. We note the followings:\\
\\
$(i)$ The radial variation of pressure and energy density profiles are found positive definite inside the star both  in higher dimensional EGB and GR. By virtue of the imposed constraint the radial pressure ($p_{r}$) vanishes at the boundary, but the tangential pressure is nonzero. The plot in Fig.(2) shows that $p_{r}$ increases for higher value of $\alpha$ at the center, after a particular crossover point (say $r=r_{0}$), the pressure decreases faster for higher values of $\alpha$. The energy density and both thus pressure decreases away from the  center of the star.  The central pressure  is found less for $\alpha <0$ compared to $\alpha >0$. It is important to mention here that string theory allow a positive $\alpha$ but the negative value we found here is of academic interest and not new in theory \cite{bcp,guo}. However, we find here that in $D=6$, a stable stellar model can be constructed with negative $\alpha$ only.\\

$(ii)$ The stellar models in EGB gravity are found to satisfy the inequalities, $\frac{d\rho}{dr} < 0$, $\frac{dp_{r}}{dr}<0$ and $\frac{dp_{t}}{dr} < 0$.\\

$(iii)$ In EGB gravity with higher dimensional FS metric, an anisotropic star of a given mass accommodates in a smaller radius for  larger positive magnitude  of  $\alpha $, or in other words, it admits a more compact star.  A compact star with  same mass as that of  the PSR J0348+0432 can be accommodated with radius $b=11 \;km. $ also in a $5$- dimensional Einstein gravity ($\alpha = 0$). In the EGB gravity, a star with radius smaller than $11$ km. can be accommodated for the same mass as  obtained in GR. But a negative value of $\alpha$ permits existence of compact object with a radius greater than $11 \; km.$ for the given mass which is plotted in Fig.(2) . As the radius of a star is not precisely known, we can predict the radius of a stable star once the mass is known for different  $\alpha$. Thus we can determine $\alpha $ if the mass and radius are known from observation. The coupling constant $\alpha$ in the EGB gravity is playing an important role in determining the physical features of the compact objects. \\

$(iv)$ The radial variation of the measure of anisotropy plotted in Fig.(3) confirmed that  $p_{t} > p_{r}$, $\it i.e.$ the anisotropic stress is directed outwards, hence there exists a repulsive force that allows the formation of super massive stars, here $\Delta$ is zero at the center which  develops a non-zero value away from the center. Thus the EGB-terms modifies the structure of compact object, perhaps it admits a compact star with a hard core that might exist near the center.\\

$(v)$ The fluid inside the star is found causal as the sound speed is subluminal  evident from  Figs.(4) and (5). Thus the stellar models obtained in EGB gravity with FS-metric accommodates a physically realistic stable stellar configuration.\\

$(vi)$ The parameter  $\Gamma$ plotted in Fig.(6) for different values of $\alpha$ indicates that $\Gamma >\frac{4}{3}$ inside the star,  although one finds $\Gamma$ for $\alpha >0$, less than that for $\alpha <0$. Thus in all the cases strange star can be accommodated.\\

$(vii)$ The plots of energy conditions, namely, weak, strong and dominant energy conditions in Figs.(7) show that the ECs satisfies inside the star.\\

$(viii)$ The profiles of the three different forces inside the star are shown for different values of $\alpha$ in Fig.(8) and found to satisfy TOV equation. The radial variation of the forces $f_{h}$ and $f_{g}$ increases in opposite direction as $\alpha$ increases, however,  the radial variation of $f_{a}$ is found less compared to the other two forces. The variation of $f_{a}$ is very small throughout the star. In the case of GR ($\alpha=0$), the three forces are comparatively lower than that  in the EGB gravity  (Fig. (8)). The equilibrium configuration of the anisotropic star is  ensured as the hydrostatic  and anisotropic forces counter balance the gravitational force since a model satisfy the TOV equation.\\

$(ix)$ From Fig.(9), it is evident that for a given dimension as the coupling parameter $\alpha$ increases the maximum mass bounded by a star of given radius increases. We note that a star with $M$ = 2.01$\pm$0.04 $M_{\odot}$ can be accommodated to a  smaller radius when $\alpha > 0$  in EGB gravity compared to a star in pure GR   $\alpha=0$. Thus more compact objects can be accommodated in EGB gravity.\\

$(x)$ In  Table-1, for $\alpha = 0$ in $D=5$-dimensions Einstein gravity it permits a star of radius $b=11.05 \, km.$. Both the central density and the radial pressure at the center of the star are found to increase with the increase in the coupling constant $\alpha$. We found a new feature in EGB gravity that  a more compact star can be realized when $\alpha$ is increased in $D=5$ dimension. Thus super dense stars can  be accommodated in the EGB gravity compared to that in the GR with  FS metric ($D >4$).\\

$(xi)$ In Table-2, the physical parameters in $D=6$ dimensional FS metric in EGB gravity for a star with its mass $ M= 2.01 \, M_{\odot}$ are tabulated. In a six dimensional EGB gravity we obtain stable stellar models with negative $\alpha$ only although both  the signs of $\alpha$  permits stable compact objects in $D=5$ dimensions. This is a new result  which is not in agreement with the string theory,  the result obtained here is of particular interest academically, as consistent  cosmological models in EGB gravity are permitted with negative $\alpha$ \cite{bcp,guo}.\\

 $(xii) $ The negative value of $\alpha$ accommodates dense star with greater central density and pressure at the center compared to that of a positive $\alpha$ in EGB gravity.
In Table-3, parameters are determined for five different known stars $e.g.$, Vel X-1, Cen X-3, EXO 1745-268, SAX J1748.9-2021 and M13 in EGB gravity for constructing physically realistic stellar models. The best-fit values of the model parameters are tabulated in Table-3. The outcome is that a stable stellar configuration can be constructed for positive as well as  negative $\alpha$ in five dimensional FS manifold. But in six dimension only $\alpha\leq 0$ permits stable stellar model.\\
 \\
 
We present a class of new relativistic solutions in EGB gravity with higher dimensional  FS metric to construct models for strange stars in  hydrostatic equilibrium. For a given mass of a star it is shown that the stellar models with different compactification factor or radius of the star can be accommodated for matter described by MIT Bag model EoS. It is noted  that a stable compact object is permitted with  positive as well as  negative values of GB coupling constant in $D =5$ but stellar  models are  permitted in $D=6$ with $\alpha \leq 0$ only.
Although the mass of a star is known precisely, the radius is not known very precisely, under this circumstance the models obtained here describes a class of stars with  different compactification. The precise measurement of the radius of a strange star in future will be helpful in understanding the role of EGB gravity.
 In conclusion the investigation carried out here  permits a class strange stars determined by the coupling parameters $\alpha$ and space-time dimensions. The GB curvature terms improves the likelihood of the stellar models conforming realistic distributions. Implication of our results, in the context of current observational data of compact stars, needs to be probed further taking into account  rotation which will be taken up elsewhere.\\

\section{Acknowledgements}
SD is thankful to UGC, New Delhi for financial support. BCP acknowledges North Bengal University with a research grant. The authors
would like to thank IUCAA Centre for Astronomy Research and Development (ICARD), NBU for extending research facilities.

\end{document}